\documentclass[aps,groupedaddress,superscriptaddress,amssymb,amsmath,amsbsy,amsfonts,twocolumn,pra]{revtex4}
\usepackage{amsmath}
\usepackage{amssymb}
\usepackage{graphicx,color}
\usepackage{amsthm}
\usepackage{hyperref}
\usepackage{extarrows}
\usepackage{adjustbox,lipsum}
\usepackage{makecell}
\usepackage{bbold}

\begin{document}

\newcommand*{\cl}[1]{{\mathcal{#1}}}
\newcommand*{\bb}[1]{{\mathbb{#1}}}
\newcommand{\ket}[1]{|#1\rangle}
\newcommand{\bra}[1]{\langle#1|}
\newcommand{\inn}[2]{\langle#1|#2\rangle}
\newcommand{\proj}[2]{| #1 \rangle\!\langle #2 |}
\newcommand*{\tn}[1]{{\textnormal{#1}}}
\newcommand*{\1}{{\mathbb{1}}}
\newcommand{\T}{\mbox{$\textnormal{Tr}$}}
\newcommand*{\todo}[1]{\textcolor[rgb]{0.99,0.1,0.3}{#1}}

\theoremstyle{plain}
\newtheorem{prop}{Proposition}
\newtheorem{proposition}{Proposition}
\newtheorem{theorem}{Theorem}
\newtheorem{lemma}[theorem]{Lemma}
\newtheorem{remark}{Remark}

\theoremstyle{definition}
\newtheorem{definition}{Definition}

\title{Entropic measure of directional emissions in microcavity lasers}
\author{Kyu-Won Park}
\affiliation{Research Institute of Mathematics, Seoul National University, Seoul 08826, Korea}
\author{Chang-Hyun Ju}
\affiliation{Department of Electronic Engineering, Yeungnam University, Gyeongsan 38541, Korea}
\author{Kabgyun Jeong}
\email{kgjeong6@snu.ac.kr}
\affiliation{Research Institute of Mathematics, Seoul National University, Seoul 08826, Korea}
\affiliation{School of Computational Sciences, Korea Institute for Advanced Study, Seoul 02455, Korea}

\date{\today}

\begin{abstract}
We propose a new notion of the directional emission in microcavity lasers. First, Shannon entropy of the far-field profiles in the polar coordinate can quantify the degree of unidirectionality of the emission, while previous notions about the unidirectionality can not efficiently measure in the robust range against a variation of the deformation parameter. Second, a divergence angle of the directional emission is defined phenomenologically in terms of full width at half maximum, and it is only easily applicable to a simple peak structure. However, Shannon entropy of semi-marginal probability of the far-field profiles in the cartesian coordinate can present equivalent results, and moreover it is applicable to even the cases with a complicated peak structure of the emission.
\end{abstract}

\maketitle

Microcavity lasers have recently garnered considerable attention owing to their applicability as optimal candidate models for studying wave chaos~\cite{cao2015dielectric,liu2018chaotic} and non-Hermitian quantum systems~\cite{cao2015dielectric,takata2022imaginary,park2020maximal} as well as their optoelectronic applications and photonics~\cite{vahala2003optical} such as speckle-free full-field imaging~\cite{redding2015optical}, broadband coupling~\cite{jiang2017optical}. In addition, recent studies reveal many interesting physics in phase space related to the photon transport~\cite{chen2019combining,qian2021combining}. In particular, the key to such applications is that microcavity lasers have high quality ($Q$) factors and directional emission simultaneously~\cite{wiersig2008combining, yan2009directional}. High-$Q$ modes guarantee a low threshold lasing; in addition, they can be adopted for bio-molecule detections~\cite{armani2007label,vollmer2008whispering} and nano particles~\cite{zhu2010chip,chen2017exceptional}. However they are limited by isotopic emissions and low output power. Hence, directional emission is also required to support a high output power, and also facilitate easy coupling to a waveguide for optoelectronic circuits. Therefore, to date, several microcavity lasers have been studied to achieve these properties simultaneously~\cite{wang2009deformed,yang2016mode,song2011highly,peter2018control,wang2018high,yang2021design}.

In these studies, the discussions about a high $Q$-factor have been addressed quantitatively and systemically, because the $Q$-factor can be well defined by $Q=\frac{f_{r}}{\Delta f}$ where $f_{r}$ is a resonance frequency and $\Delta f$ is a resonance width, or equivalently, $Q=\frac{f_{r}}{2|f_{i}|}$ in the context of a complex eigenfrequency $f_{c}=f_{r}+jf_{i}$. However, regarding directional emission, we consider that its definition is relatively subtle and it remains not well established up to date. In this paper, thus, we introduce new measures of the unidirectional emission by exploiting the notion of Shannon entropy, and this suggestion holds that an entropic measure of the unidirectional emission is more accurate and efficient than former alternatives related to the emission window~\cite{ryu2011designing,ryu2019optimization}. It is possible to validate these results by demonstrating that, former alternatives fail to detect the degree of the directional emission, our new methods can.

\begin{figure}
\centering
\includegraphics[width=\linewidth]{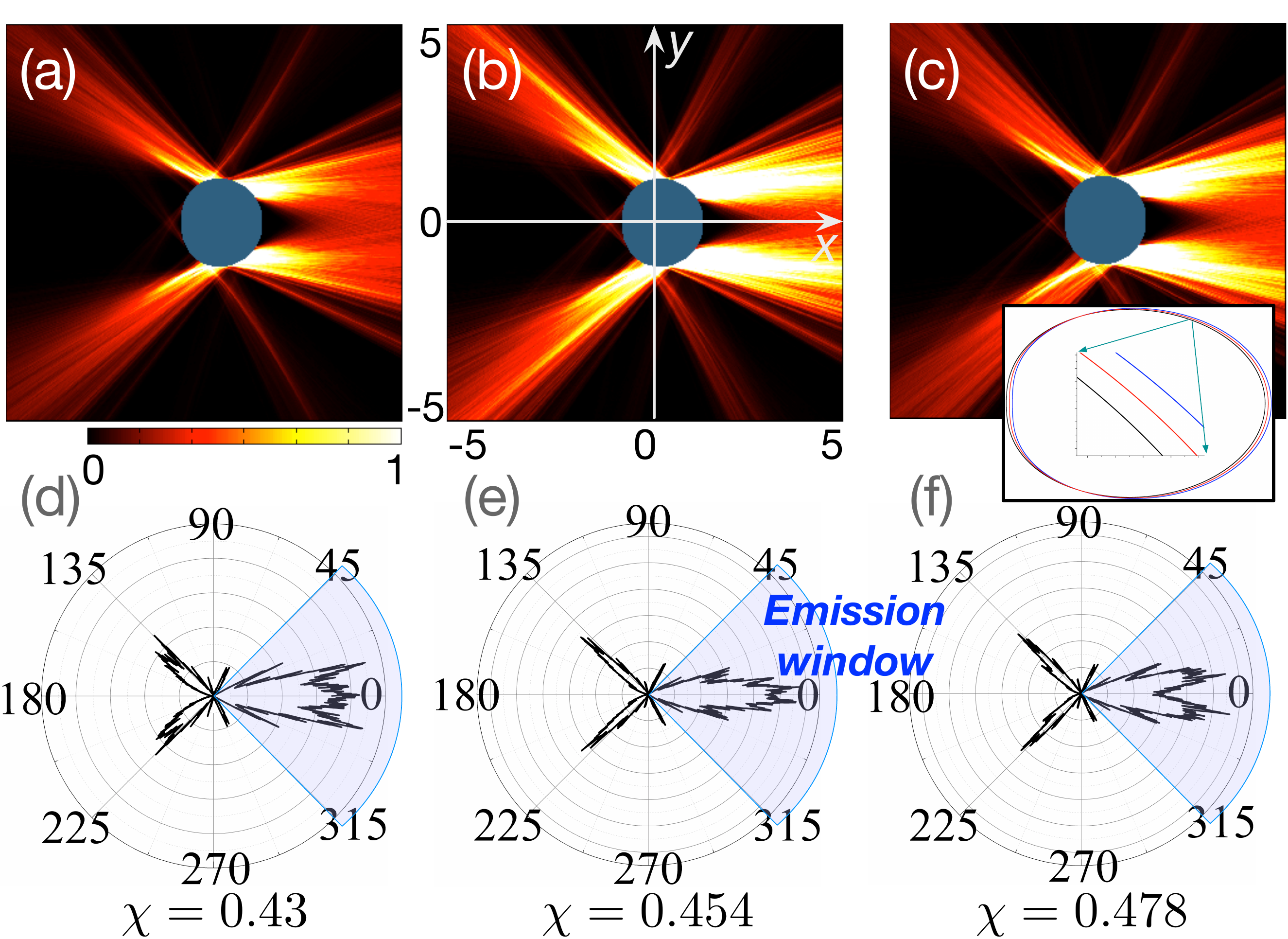}
\caption{Far-field profiles in a lima\c{c}on-shaped cavity. The figures (a), (b), and (c) are the far-field profiles in the cartesian coordinate within $x\in[-5,5]$ and $y\in[-5,5]$ at each deformation $\chi=0.43$, $\chi=0.454$, and $\chi=0.478$, respectively. The figures (d), (e), and (f) are also far-field profiles plotted in the polar coordinate corresponding to the far-field profiles in the cartesian coordinate. The angles in the range of $|\theta|\le\frac{\pi}{4}$ is the emission window. The inset in (c) shows the boundary shape of cavity at each $\chi=0.43$ (black lower line), 0.454 (red center line), and 0.478 (blue upper line), respectively.}
\label{fig:1}
\end{figure}

To further engage this discussion, we consider a prevalent and basic lima\c{c}on-shaped microcavity laser as a candidate for the directional emission with a high $Q$-factor~\cite{wiersig2008combining,yan2009directional,shinohara2009ray,song2009chaotic}. In this reason, it has been extensively studied to date. The geometrical boundary of the lima\c{c}on-shaped microcavity is defined as follows:

\begin{align}
R(\theta)=R_{0}(1+\chi\cos\theta),
\end{align}
where $\theta$ is the angle in the polar coordinate, $\chi$ is the deformation parameter, and $R_{0}(=1)$ is the radius of circles at $\chi=0$. The lima\c{c}on-shaped microcavity laser with an effective index of refraction $n=3.3$ (for InGaAsP semiconductor microcavity) are treated in ray simulation. Some of the representative far-field profiles (FFPs) in the lima\c{c}on-shaped cavity is shown in Fig.~\ref{fig:1}. The FFPs are obtained from transmitted rays by using Fresnel equations for transverse magnetic (TM) modes. The figures (a), (b), and (c) in Fig.~\ref{fig:1} are plotted in the cartesian coordinate within the range of $|x|\le5$ and $|y|\le5$ at each deformation $\chi=0.43$, $\chi=0.454$, and $\chi=0.478$, respectively. The subfigures (d), (e), and (f) in Fig.~\ref{fig:1} are FFPs plotted in the polar coordinate, corresponding to FFPs in the cartesian coordinate, respectively. We call the angles in the range of  $|\theta|\le\frac{\pi}{4}$ as `emission window', which was first introduced by Refs.~\cite{ryu2011designing,ryu2019optimization} for the definition of unidirectionality.

It was reported that the deformation parameter $\chi=0.43$ is the optimal value for unidirectional emission~\cite{wiersig2008combining}. Moreover, they have concluded that the results are robust against any variation of the deformation parameter in the range of $0.43\leq\chi \leq 0.49$~\cite{wiersig2008combining}. Consistent with their results, the overall profiles of all figures in Fig.~\ref{fig:1} appear similar to each other. However, if we observe the morphologies of FFPs for Fig.~\ref{fig:1}(d), (e), and (f) simultaneously, we can suggest that FFPs at $\chi=0.454$ exhibit a lager unidirectionality than the others.

\begin{figure}
\centering
\includegraphics[width=\linewidth]{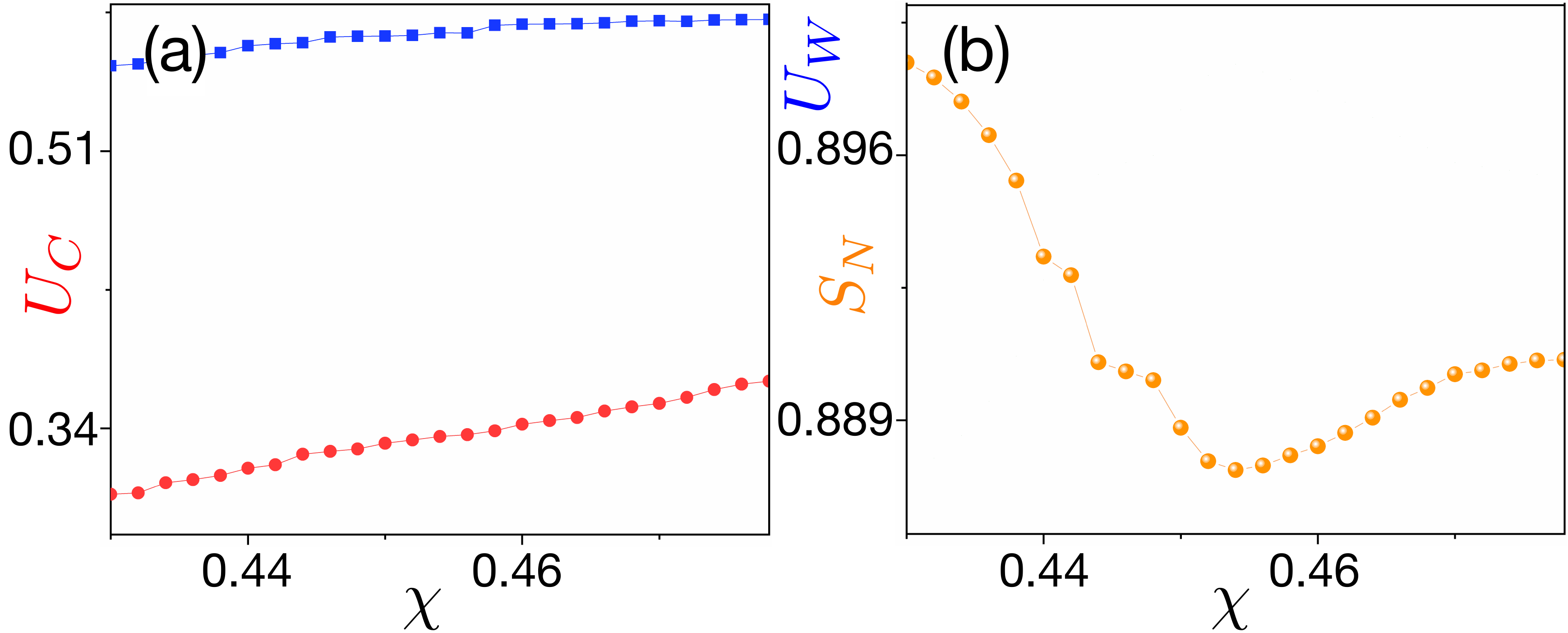}
\caption {Three types of measure of unidirectionality in the lima\c{c}on-shaped cavity. The blue squares ($U_{W}$) and red circles ($U_{C}$) in (a) are measures of unidirectionality associated with emission window. They almost increase linearly as a function of the deformation. The orange circles in (b) are normalized Shannon entropies and they have a local minimum at $\chi=0.454$.}
\label{fig:2}
\end{figure}

To validate this suggestion, we address three types of measure for the unidirectionality.  The blue squares ($U_{W}$) and red circles ($U_{C}$) in Fig.~\ref{fig:2}(a) are measures of the unidirectionality associated with the emission windows.
More precisely, the unidirectionality $U_{C}$ marked by red circles is defined as follow~\cite{shim2013adiabatic,song2010directional}:
\begin{align}
U_{C}=\frac{\int^{2\pi}_{0}I(\theta)\cos\theta d\theta}{\int^{2\pi}_{0} I(\theta)d\theta}.
\label{ep:2}
\end{align}
Here, $I(\theta)$ denotes the intensity of angular distribution of FFPs~\cite{jiang2016whispering}. The $\cos\theta$ as a window function determines the extent to which the emission directionality deviates from unidirectionality. Actually, the positive and negative $U_{C}$ represent tendencies toward a forward and backward emission, respectively, and $U_{C}=0$ corresponds to the bidirectional or isotropic emission of the microcavity laser.
In our case, the values of  $U_{C}$ almost increase linearly from $U_{C}\simeq 0.3$ to $U_{C}\simeq 0.35$ in the range of $0.43 \leq\chi \leq0.478$.
This result implies that the unidirectionality increases as the deformation parameter $\chi$ increases, unlike observing the morphologies of FFPs for Fig.~\ref{fig:1}(d), (e), and (f).
Note that we here discretize $2\pi$ into $3600$ pieces for numerical calculation, i.e., $d\theta\sim\Delta\theta=0.1^\circ$.

The other measure of the unidirectionality $U_{W}$ marked by blue squares is also defined as follow~\cite{ryu2011designing,ryu2019optimization}:
\begin{align}
U_{W}=\frac{\int^{\frac{\pi}{4}}_{\frac{-\pi}{4}} I(\theta)d\theta}{\int^{2\pi}_{0} I(\theta)d\theta},
\label{ep:3}
\end{align}
where the integral range of numerator runs from $-\frac{\pi}{4}$ to $\frac{\pi}{4}$. The angles in this range of $|\theta|\le\frac{\pi}{4}$ is so-called the emission window as mentioned before.

Hence, the meaning of this definition is clearly ratio between accumulated intensities of FFPs within emission window and the total intensity of FFPs.
The values of $U_{W}$ also increase almost linearly from $U_{W}\simeq 0.56$ to $U_{W}\simeq 0.59$ in the range of $0.43 \leq\chi \leq0.478$. Consequently, these values of $U_{W}$ are also inconsistent with our observation. We can conjecture that this discrepancy is attributed to the fact that $U_{W}$ increases proportionally to the integral value of the emission window, regardless of the detailed structure of the emission window. Furthermore, it should be noticed that $U_{C}$ exhibits a similar trend with $U_{W}$.

Next, we introduce entropic unidirectionality. Accordingly, we first have to obtain a Shannon entropy associated with FFPs. Shannon entropy is a relevant measure of the average amount of information for a random variable with a given probability distribution function~\cite{shannon1948mathematical}. It was firstly developed and utilized in communication theory~\cite{shannon1948mathematical}. However, recently it has been also exploited in various areas such as bio-system~\cite{fuhrman2000application}, economics~\cite{eagle2010network}, atomic physics~\cite{gonzalez2003shannon}, and microcavity laser~\cite{park2018shannon}. 

The discrete Shannon entropy for the intensity of FFPs is formally defined as:
\begin{align}
S_{N}=-\frac{1}{\log K}\sum^{K}_{i=1}\rho_{i}\log \rho_{i},
\end{align}
 where $\rho_{i}$ represents the probability distribution obtained under the normalization condition $\sum^{K}_{i=1} I(\theta_{i})=1$. That is, the random variable $X$ is an angular coordinate ($\Theta$) with the probability distribution $\{\rho_{i}\}=\{P(\Theta=\theta_{i})\}$. In addition, the $\frac{1}{\log K}$ is a normalization factor such that the value of Shannon entropy is restricted in the range of $0\leq S_{N}\leq 1$, with $K=3600$ as mentioned above. The orange circles in Fig.~2 (b) represent normalized Shannon entropy ($S_{N}$) calculated by this definition. The value of $S_{N}$ has a local maximum ($0.898$) at $\chi=0.43$ and local minimum ($0.888$) at $\chi=0.454$, respectively.

Our recent works have confirmed that Shannon entropy can be beneficial in measuring the delocalization of the given probability distributions~\cite{park2021indicators}. Hence, the minimum (or maximum) value of the Shannon entropy indicates the maximum localization (or delocalization) of the intensity of FFPs in the polar coordinate under the normalization condition $\sum^{K}_{i=1} I(\theta_{i})=1$. Consequently, the local minimum value of $S_N$ at $\chi=0.454$ directly verifies the maximal undirectionality. Moreover, the maximal entropy, $\log K$, corresponds to the isotropic distribution that is not a bidirectional emission in the polar coordinate. This result validates the conjecture that by exploiting the Shannon entropy, our measure of unidirectionality is coincident with our observations in Fig.~\ref{fig:1}. 

\begin{figure}
\centering
\includegraphics[width=\linewidth]{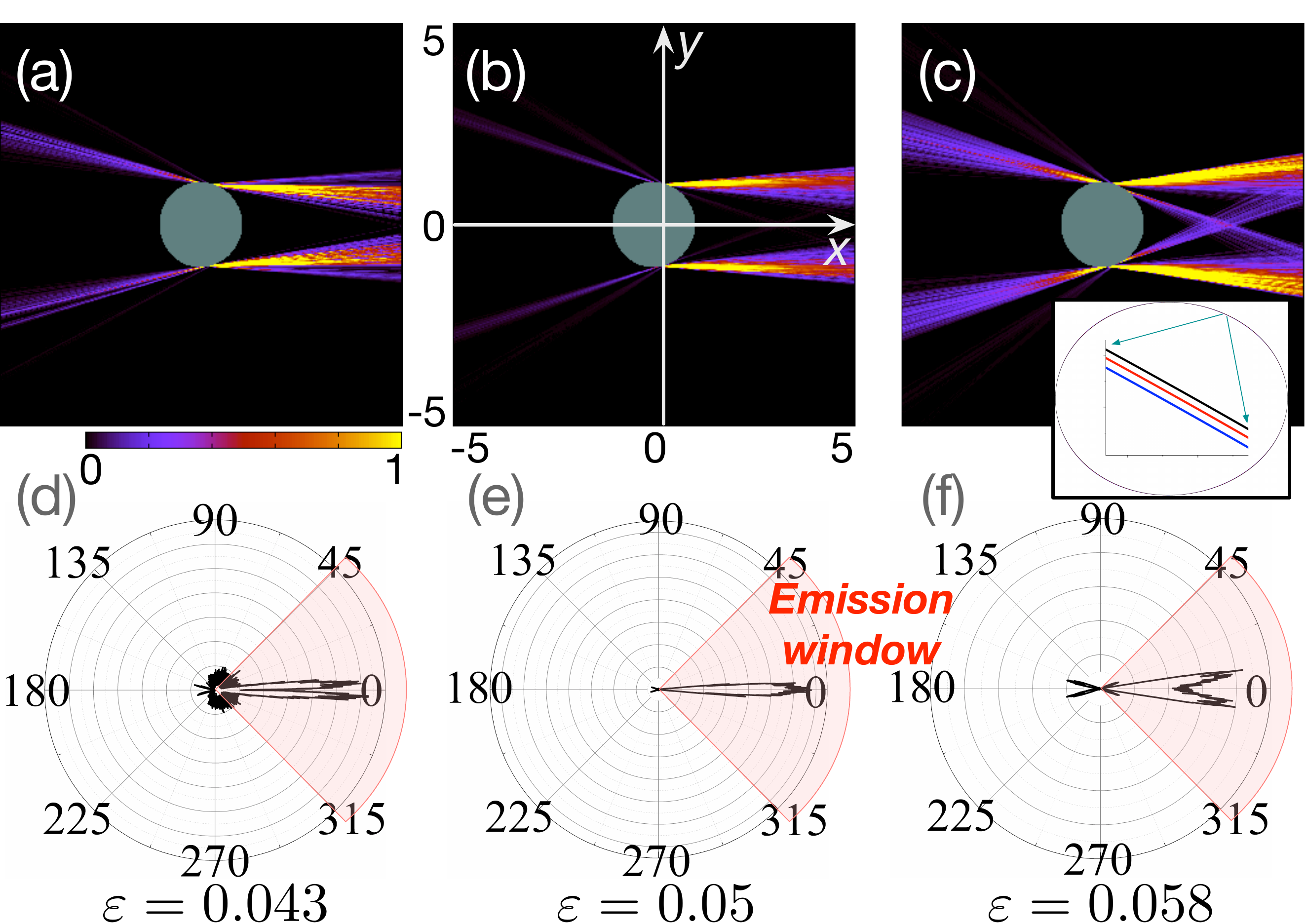}
\caption {FFPs in an oval-shaped microcavity laser. The subfigures (a), (b), and (c) are FFPs in the cartesian coordinate restricted in the range of $x\in[-5,5]$ and $y\in[-5,5]$ at each deformation $\varepsilon=0.043$, $\varepsilon=0.05$, and $\varepsilon=0.058$, respectively. The subfigures (d), (e), and (f) are also FFPs plotted in the polar coordinate corresponding to FFPs in the cartesian coordinate. The inset in (c) shows the boundary shape of cavity at each $\varepsilon=0.043$ (black upper line), 0.05 (red center line), and 0.058 (blue lower line), respectively.}
\label{fig:3}
\end{figure}

For more generality of our argument and to understand the discrepancy between Fig.~\ref{fig:2}(a) and Fig.~\ref{fig:2}(b), let us consider an oval-shaped microcavity laser~\cite{lee2018extremely}. The FFPs are also obtained from transmitted rays by using Fresnel equations for transverse electric (TE) modes with an effective index of refraction $n=3.3$. The geometrical boundary condition of an oval shaped-cavity, which is the deformed from an ellipse, is defined as follow:
\begin{align}
\frac{x^2}{a^2}+(1+\varepsilon x)\frac{y^2}{b^2}=1.
\end{align}
For convenience, we substitute the deformation parameter $\chi$ to $\varepsilon$.
It was reported that optimized condition for a directional emission is $a=1.0$, $b=1.03$, and $\varepsilon=0.05$ where $a$ and $b$ are major and minor axis of an ellipse, and $\varepsilon$ is the deformation parameter, respectively~\cite{lee2018extremely}. According to their results, we conduct ray simulation in the range of $ 0.043\leq\varepsilon\leq 0.058$ at the fixed value of $a=1.0$ and $b=1.03$. We plot some of the representative FFPs, i.e., Fig.~\ref{fig:3}(a), (b), and (c) in the cartesian coordinate restricted in $x\in[-5,5]$ and $y\in[-5,5]$ at each deformation $\varepsilon=0.043$, $\varepsilon=0.05$, and $\varepsilon=0.058$. The corresponding FFPs in the polar coordinate are displayed in Fig.~\ref{fig:3}(d), (e), and (f).

\begin{figure}
\centering
\includegraphics[width=\linewidth]{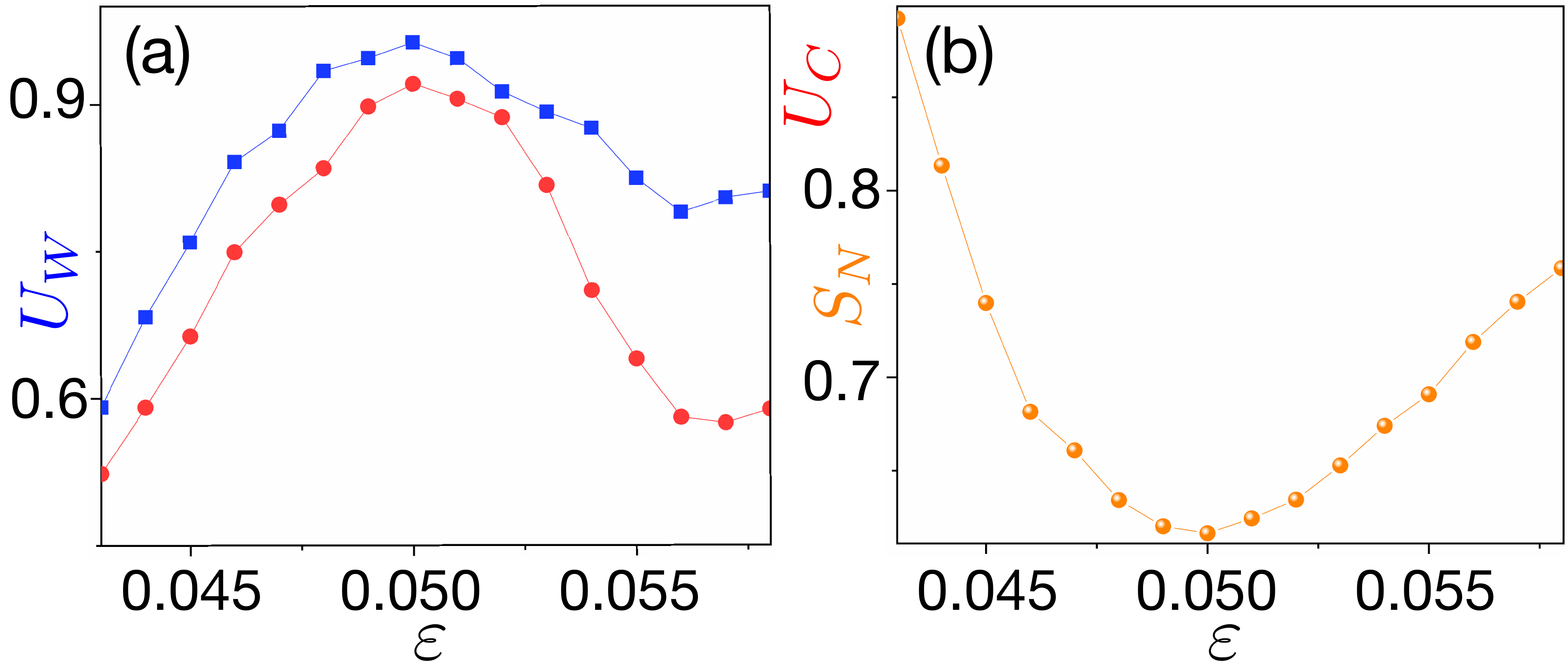}
\caption {Three types of measure of the unidirectionality in the oval-shaped microcavity.  The blue squares and red circles in (a) are measures of the unidirectionality associated with the emission window. Both of them have local maximal values at $\varepsilon=0.05$. The orange circles in (b) are normalized Shannon entropies. The values of orange circles have a local minimum value at $\varepsilon=0.05$.}
\label{fig:4}
\end{figure}

On the contrast to Fig.~\ref{fig:1}, we can easily notice that the overall FFPs depending on the deformation parameter vary manifestly, and it can be naturally expected that the local maximum of the unidirectionality is obtained at $\varepsilon=0.05$ by observing Fig.~\ref{fig:3}(b) and (e), and comparing other subfigures in Fig.~\ref{fig:3}. In this case, the local minimum value of $S_N$, and local maximum values of $U_W$ and $U_C$ are attained at $\varepsilon=0.05$ simultaneously, i.e., entropic measure of the unidirectionality $S_N$ agrees well with the the former measure of the unidirectionality ($U_{C},U_{W}$) related to the emission window. Consequently, this fact implies that $U_{C}$ and $U_{W}$ only can detect the unidirectionality when the overall FFPs vary significantly with a manifest variation of the emission window. However, our new measure for the unidirectionality by employing the Shannon entropy can efficiently capture the unidirectionality in any case. Notice that the relative difference between the local maximum and the local minimum of $S_N$ in Fig.~\ref{fig:4} is much larger than that of Fig.~\ref{fig:2}. This fact coincides with our initial intuition.

Another aspect for definition of the directional emission along with the unidirectionality is a so-called `divergence angle'. This concept presents an analogy for the full width at half maximum (FWHM)~\cite{yan2009directional}. In our case, the divergence angle estimates the width of the peak emission (i.e., lobe around $\theta=0$)~\cite{yan2009directional,wang2009deformed} of the emission intensity in the polar coordinate, and this can indicate the degree of the spread of FFPs onto $y$-axis in the cartesian coordinate. Under these assumptions, we fist introduce the concept of semi-marginal probability distribution. When the joint probability distribution function of random variables $X$ and $Y$ is given by $\rho(x,y)$, the marginal probability distribution of $Y$ is $\rho(y)=\int \rho(x,y)dx$ where integral is carried out over all points in the range of $(X,Y)$ for which $Y=y$. We can interpret the intensity of FFPs, $I(x,y)$, in the cartesian coordinate as a joint probability distribution function under the normalization condition $\int\int I(x,y) dxdy=1$, i.e.,  the random variables $X$ and $Y$ are components of the cartesian coordinate with the joint probability distribution function $\rho(x,y)={P(X=x, Y=y)}$. The integral $\int \rho(x,y)dx$ is performed over the interval $x\in[0, 5]$ to solely handle the forward emission. This is why we call it as semi-marginal probability distribution. Note that we have discretized ($x$,$y$)-coordinate in the range $x\in[-5,5]$ and $y\in[-5,5]$ into $1000\times1000$ grid for the numerical calculation. Then, the discrete Shannon entropy from the $\rho(y_{j})$ is defined by
\begin{align}
S_{y}=-\frac{1}{\log K}\sum^{K}_{j=1}\rho_{y_{j}}\log \rho_{y_{j}},
\end{align}
where the semi-marginal distribution $\rho_{y_{j}}=\sum^{K}_{i=1} \rho(x_{i},y_{j})$ and $K=1000$.

\begin{figure}
\centering
\includegraphics[width=7cm]{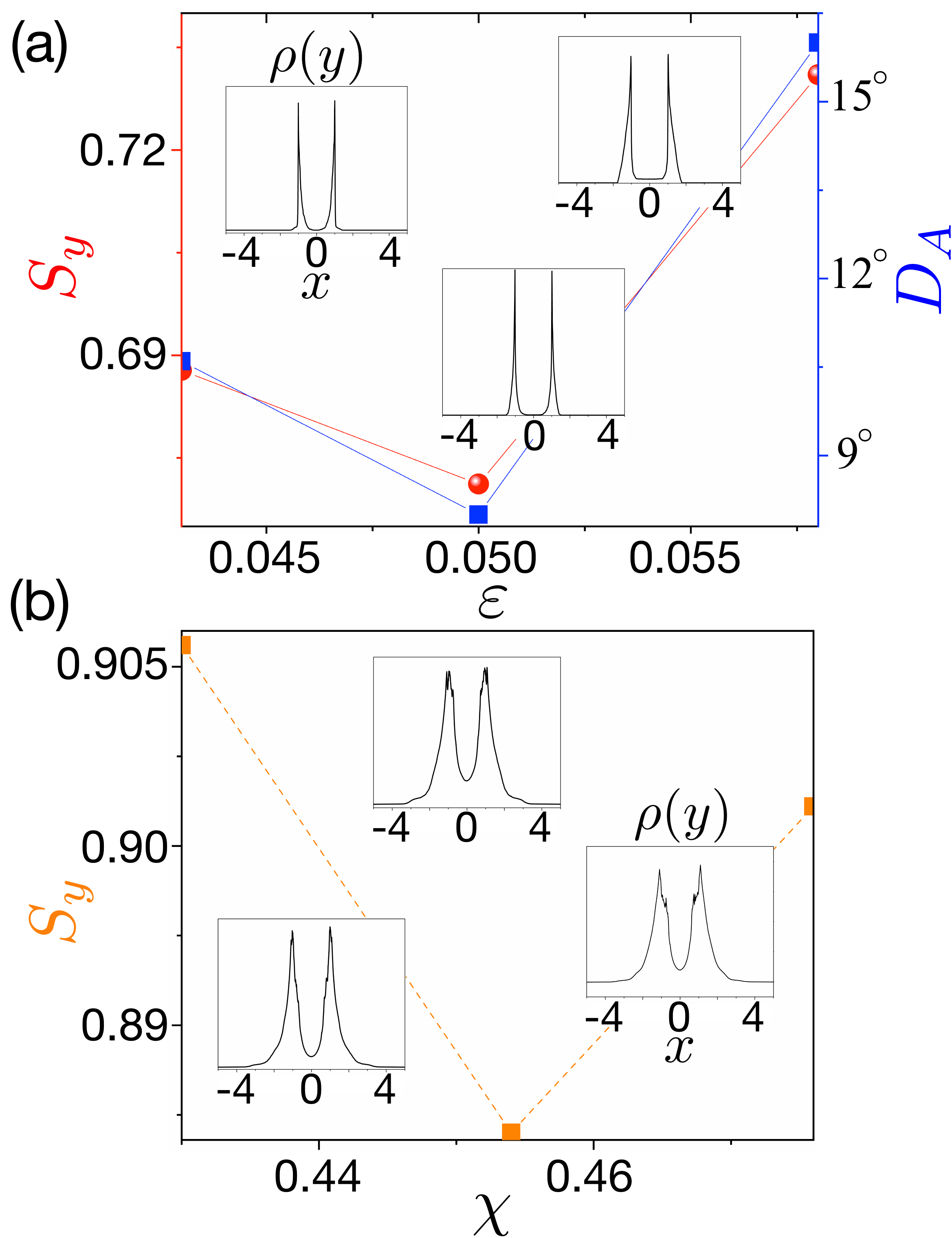}
\caption {(a) The blue squares are markers of the divergence angle ($D_A$) and the red circles are those of Shannon entropy of the semi-marginal probability densities ($S_{y}$) in the oval-shaped microcavity laser at each  $\varepsilon=0.043$, $\varepsilon=0.05$, and $\varepsilon=0.058$. These two plots show similar trends. Three insets are semi-marginal probabilities related to $S_{y}$. (b) The orange squares are markers of Shannon entropy of semi-marginal probabilities ($S_{y}$) in the lima\c{c}on-shaped microcavity laser at $\chi=0.43$, $\chi=0.454$, and $\chi=0.478$. Three insets are also semi-marginal probabilities related to $S_{y}$.}
\label{fig:5}
\end{figure}

The blue squares in Fig.~\ref{fig:5}(a) represent the divergence angle ($D_A$) and the red spheres in Fig.~\ref{fig:5}(a) represent the Shannon entropy ($S_{y}$) of the semi-marginal probabilities in the oval-shaped microcavity laser at each $\varepsilon=0.043$, $\varepsilon=0.05$, and $\varepsilon=0.058$. Note that these two plots show similar trends and this results support our previous assumption. Three insets in Fig.~\ref{fig:5} represent semi-marginal probability densities ($\rho_{y}$). All peaks of $\rho_{y}$ are located at $\varepsilon\simeq \pm 1$, and careful examination reveals that the third one is more spread out than the others.

In the case of complex peak structures, as illustrated in Fig.~\ref{fig:1}(d), (e), and (f), we can barely define $D_A$ (i.e., FWHM). However, we can quantitatively and systematically  measure the spread of the emission peak by exploiting $S_{y}$, and the obtained results are presented in Fig.~\ref{fig:5}(b).
The orange squares in Fig.~\ref{fig:5}(b) indicate Shannon entropy ($S_{y}$) of semi-marginal probability ($\rho_{y}$) in the lima\c{c}on shaped-cavity at $\chi=0.43$, $\chi=0.454$, and $\chi=0.478$. The minimum value of $S_{y}$ in Fig.~\ref{fig:5}(b) is substantially larger than that of $S_{y}$ in Fig.~\ref{fig:5}(a), which can be confirmed by comparison between Fig.~\ref{fig:1}(e) and Fig.~\ref{fig:3}(e).

Consequently, we can say that the oval-shaped micro-cavity laser has larger unidirectionality and smaller $S_y$ (i.e., $D_A$) than the lima\c{c}on-shaped microcavity laser at least in our examples. In contrast, the lima\c{c}on-shaped microcavity laser has more robust range against any variation of the deformation parameter than the oval-shaped microcavity laser.

We have present new measures for a directional emission in microcavity lasers by exploiting the Shannon entropy. There are primarily two aspects of the directional emission---the unidirectionality and the divergence angle. Shannon entropy obtained from the normalized intensity of angular distributions of FFPs can measure the unidirectionality even if former notions can not effectively detect the directional emission when the emission is robust against a variation of the deformation parameter. However, it should be noticed that our method can be wrong in some extreme cases. For example, extreme bidirectional emission can yields a very low value of the Shannon entropy. In order to exclude such cases, each FFPs must have a single broad but dominant emission. To guarantee this condition, an overlap between FFPs and Gaussian normal distribution must be larger than a critical value as the role of auxiliary measure. In the proposed examples, the overlaps are always larger than 0.6 with a mean value $\langle\theta\rangle=0$ (corresponding to the emission peak) and standard deviation $\sigma=45^{\circ}$ (corresponding to the emission window). Shannon entropy obtained from the semi-marginal probability densities of FFPs in the cartesian coordinates can provide equivalent results to the divergence angle; moreover, it can be applied to even in the case of complicated peak structures. 

Our new measure is entirely defined by normalized FFPs (as probability distributions), regardless of the origin of emissions. Thus, our measure is applicable to the FFPs obtained from the wave simulations and, furthermore, it can be applicable to any shape of microcavity lasers and, various antenna structures. We hope that our results can help to design and to modulate a micocavity lasers for a better directional emission.

\section*{Acknowledgments}
This work was supported by the National Research Foundation of Korea, a grant funded by the Ministry of Education (Grant Nos. NRF-2021R1I1A1A01052331 \& NRF- 2021R1I1A1A01042199) and the Ministry of Science and ICT (Grant No. NRF-2020M3E4A1077861).


%

\end{document}